\documentclass[final,5p,times,two-column,authoryear]{elsarticle}
\usepackage{amssymb}
\usepackage{lipsum}
\usepackage{float}
\usepackage{lineno}
\usepackage{graphicx}
\graphicspath{ {./Figures/} }
\usepackage{multirow}
\usepackage{array}
\usepackage{makecell}
\usepackage{array}
\usepackage{tabularx}
\usepackage{natbib}
\usepackage{setspace}
\usepackage{multicol}
\usepackage{lineno}
\usepackage{amsmath}
\usepackage[font=small,labelfont=bf,labelsep=space]{caption} 
\usepackage{xcolor}
\setcitestyle{square,numbers}
\begin{document}
\begin{frontmatter}

\title{PLASMA-METAL JUNCTION: A JUNCTION WITH NEGATIVE TURN-ON VOLTAGE}

\author[add1]{Sneha Latha Kommuguri}	
\affiliation{organization={Department of Physics, Pondicherry University},
addressline={Kalapet}, city={Puducherry}, postcode={605014}, state={Puducherry},
country={India}}, \author[add1]{Smrutishree Pratihary}, \author[add1]{Thangjam Rishikanta Singh}	\author[add1]{Suraj Kumar Sinha}
\
\begin{abstract}
Unlike junctions in solid-state devices, a plasma-metal junction (pm-junction) is a junction of classical and quantum electrons. The plasma electrons are Maxwellian in nature, while metal electrons obey the Fermi-Dirac distribution. In this experiment, the current-voltage characteristics of solid-state devices that form homo or hetero-junction are compared to the pm-junction. Observation shows that the turn-on voltage for pn-junction is $0.5V$ and decreases to $0.24V$ for metal-semiconductor junction. However, the pm-junction's turn-on voltage was lowered to a negative value of $-7.0V$. The devices with negative turn-on voltage are suitable for high-frequency operations.  Further, observations show that the current-voltage characteristics of the pm-junction depend on the metal's work function, and the turn-on voltage remains unchanged. This result validates the applicability of the energy-band model for the pm-junction. We present a perspective metal-oxide-plasma (MOP), a gaseous electronic device, as an alternative to metal-oxide-semiconductor (MOS), based on the new basic understanding developed. 
\end{abstract}
\begin{highlights}
\item plasma-metal junction (pm-junction) is a junction of Maxwellian plasma electrons and Fermi-electrons in the metal.
\item Comparative study of I-V Characteristics of solid state device (SSD) to a gaseous electronic device (GED).
\item Observed negative turn-on voltage for the plasma-metal junction.
\item Demonstrates the applicability of energy band theory for pm-junction. 
\item Opens up possible applications for the plasma-metal junction in high-frequency oscillators and amplifiers.
\item Presents perspective structure of a metal-oxide-plasma (MOP)  structure analogous to metal-oxide-semiconductor (MOS) device.
\end{highlights}


\end{frontmatter}


\section{Introduction}
\label{introduction}
The success of solid-state devices (SSDs) owes to the fundamental understanding of homo and hetero junctions based on the energy-band model \cite{neeman2007semiconductor}. Several vacuum and gaseous electronics (i.e., plasma) based technologies were invented in the 1950s \cite{jones1955gaseous,loeb2023basic, hirsh2012gaseous}, however, the advent of SSDs subsided and or substituted them within 20 years. Nevertheless, the new computing concepts for faster processing, such as quantum computing \cite{knill2010quantum} and chaos computing \cite{shinbrot1993using,murali2003realization} have made researchers look beyond SSDs. Interestingly, at the same time, gaseous electronics are once again leading new emergent areas such as plasma-biotechnology \cite{bovzena2018non}, plasma
medicine \cite{weltmann2016plasma,kong2009plasma}, multiphase plasmas \cite{ishijima2013high}, environmental applications \cite{mumtaz2023review,hammer1999application}, and atmospheric-pressure plasma systems \cite{kogelschatz2004atmospheric}. It has improved the overall physics and engineering aspects of plasma-based devices. The proposed concept for the realization of chaos computing \cite{murali2003realization}, the experimental observation of order-chaos-order transition of plasma fluctuations\cite {alex2015order, alex2020coexistence, jayaprakash2021doubly}, and the interplay between the coupled state of ion and electron plasma waves \cite {alex2015order, alex2020coexistence, jayaprakash2021doubly, singh2023interplay} presents the possibility of realization of chaos computing using gaseous electronics, i.e., plasmas. Therefore, pm-junction can be treated as a basic building block of gaseous electronics devices (GEDs), as pn-junction is for SSDs. In this work, we compare the I-V characteristics of a pn-junction and Schottky barrier diode (i.e., SSDs) to a pm-junction (i.e., a GED).
\par
In electronics, the small applied voltage in the range of $\sim 1-10V$ is significant in most of the applications, and small potential barriers developed across a junction of different materials play a critical role in the flow of current. Contrary to this, despite knowing that the plasma-material interface defines plasma characteristics and discharge characteristics such effects are generally ignored \cite{rapp2013development,hess1990plasma}. There is a need for accurate measurement in low voltage ranges, with a gaseous electronics point of view, for technological applications such as plasma display \cite{boeuf2003plasma} and plasma actuators \cite{roth2003aerodynamic}. However, no study reported that addresses these issues with a gaseous electronics point of view until recently a new concept of pm-junction was introduced \cite {arumugam2020plasma}.
\par
A pm-junction is an abnormal junction as the electron densities and associated parameters vary abruptly across the junction \cite{bonitz2019towards}. In most of the experimental conditions, the electron density of glow discharge plasma is about $10^{16}$$m^{-3}$ \cite{chapmanbrian}, and the electron density in metals is about $10^{28}m^{-3}$ \cite{neeman2007semiconductor}. Accordingly, it is a junction of Maxwellian plasma electrons \cite{lieberman1994principles} on one side of the junction and the metal electrons obeying Fermi-Dirac distribution \cite{neeman2007semiconductor} on the other. The Fermi-energy level for plasmas \cite {arumugam2020plasma} was defined, enabling the application of the energy-band diagram model for pm-junction. Interestingly, this concept has been further extended successfully for plasma-solution junction \cite{niu2022plasma} and a quantum mechanical description for pm-junction has also been reported recently \cite{muthu2023On}. These developments suggested the possible use of GEDs to overcome the limitations of SSDs. 
\par
The success of SSDs is based on an accurate understanding of the governing physics of semiconductor junctions based on the energy-band diagram model \cite{neudeck1983pn,dimitrijev2012principles} that explains the I-V characteristics of homo and hetero junctions. Similarly, understanding the pm-junction is critically important for advancing gaseous electronics for new applications. Therefore, we aim to simplify and accurately understand a pm-junction. Accordingly, we studied the I-V characteristics of a pm-junction which is a GED, and compared it to pn-junction, and metal-semiconductor junction the SSDs having different features of the potential barrier.
\par
In the following Sec.2, the basic concepts of pn-junction, metal-semiconductor junction, and pm-junction are briefly revisited. The experimental details for I-V characteristics measurement are given in Sec.3. The experimental observation for pm-junction for 0.3 mbar pressure is discussed in Sec. 4. The obtained I-V characteristics are compared and analysed in Sec. 5 and the conclusion of the present study is given in the last Sec.6.
\section{Theory} 
In this section, the basics of junction physics are revisited briefly for comparison with the characteristics of pm-junction. 
\subsection{pn-junction and metal-semiconductor junction}
	The theory of pn-junction and metal-semiconductor junction is well established and explains correctly the observed I-V characteristics. Understanding the I-V characteristics is crucial for designing circuits involving diodes, rectifiers, voltage regulators, etc. The energy band diagram is a graphical representation to explain the I-V characteristics of the diode \cite{neeman2007semiconductor}. Real-world diodes vary slightly from ideal diodes due to factors like temperature and impurities in the material.  The current-voltage relationship for the pn-junction is given by
       \begin{equation}
	I=\left[\frac{eD_pp_{n0}}{L_p}+\frac{eD_nn_{p0}}{L_n}\right]\left[exp\left(\frac{eV_a}{kT}\right)-1\right]
	\end{equation} 
        \begin{equation}
	I=I_s\left[exp\left(\frac{eV_a}{kT})\right)-1\right]
	\end{equation}
    The reverse saturation current, $I_s$ is defined as
        \begin{equation}
	I_s=\left[\frac{eD_pp_{n0}}{L_p}+\frac{eD_nn_{p0}}{L_n}\right]
	\end{equation}
 $D_p$ and $D_n$ are the diffusion coefficients of holes and electrons.\\
 $p_{n0}$ is the thermal equilibrium minority carrier electron concentration in the p region, and $n_{p0}$ is the thermal equilibrium minority carrier hole concentration in the n region. $L_n$ and $L_p$ are the minority carrier electron and hole diffusion lengths. The reverse saturation current relies on the diffusion and drift of minority carriers.
 At thermal equilibrium, there is no net current flow across the junction due to the drift of the carriers in the electric field that cancels the diffusion current \cite{streetman2000solid}. As the voltage increases, more charge carriers are injected across the junction, causing a rapid increase in the current. The exponential term in eq (2) accounts for an increase in the carrier concentration with an increase in the applied voltage ($V_a$) across the junction.
\par
The I-V characteristics of a metal-semiconductor junction differ from the standard pn-junction due to its junction formation of metal and semiconductor. In forward bias, the metal-semiconductor junction behaves exactly as a pn-junction but has a lower turn-on voltage. This reduced turn-on voltage is due to the lower barrier height \cite{neeman2007semiconductor, streetman2000solid}. The I-V characteristics of metal-semiconductor junction is given by
\begin{equation}
	I=\left[A^*T^2exp\left(\frac{-e\phi_{Bn}}{kT}\right)\right]\left[exp\left(\frac{eV_a}{kT}\right)-1\right]
	\end{equation}
         \begin{equation}
	I=I_{sT} \left[exp\left(\frac{eV_a}{kT}\right)-1\right]
	\end{equation}
Where $I_{sT}$ is the reverse saturation current and is given by
\begin{equation}
I_{sT}=\left[A^*T^2exp\left(\frac{-e\phi_{Bn}}{kT}\right)\right]   
\end{equation}
The parameter $A^*$ is the effective Richardson constant for thermionic emission, and $\phi_{Bn}$ is the Schottky barrier height. The exponential term in the reverse saturation current represents the probability of electron tunnelling across the barrier. As the barrier height ($\phi_B$) increases, the exponential term decreases, showing a lower probability of electron tunnelling through the barrier and vice-versa \cite{neeman2007semiconductor}. Therefore, the significant difference between the pn-junction and the metal-semiconductor junction is the turn-on voltages, the function of the barrier height of the metal-semiconductor junction, and the doping concentrations of the pn-junction. This leads to differences in the frequency response or switching characteristics\cite{neeman2007semiconductor, streetman2000solid}.
        
\subsection{pm-junction}
The pn-junction is a homo-junction of the same semiconducting material (i.e., Silicon) with different doping, and the metal-semiconductor junction is a hetero-junction. However, the pm-junction is a junction of two different states of matter, i.e., the solid metals and ionised gas as plasma. 
The I-V Characteristic of a pm-junction is \cite{hershkowitz1989langmuir} is expressed as 
        \begin{equation}
	I=I_{es} \left[exp\frac{\left(V-V_P\right)}{T_e} \right]
	\end{equation}
where $V_P$ is the plasma potential,  $T_e$ is the electron temperature, $I_{es}$ is the electron saturation current, which is given by
\begin{equation}
 I_{es}=\frac{1}{4}enAu_e   
\end{equation}
where $u_e$ is the electron's average velocity, $n$ is the plasma density, and $A$ is the area of the probe. The above expression does not include material property. This suggests that even if we use Stainless steel or tungsten, the current across the junction will remain unaffected. However, in the experiments, it is otherwise. This limitation was addressed by the energy-band model for plasma-metal junction\cite{arumugam2020plasma}.
\par
At the interface of plasma and metal, there is a potential barrier due to the different work functions of metal and the plasma potential \cite{arumugam2020plasma}. Taking all these into consideration and with the definition of Fermi energy level for plasma, the energy-band diagram for a pm-junction \cite {arumugam2020plasma} introduces a correction and accounts for the role of the work function of the metal ($\phi_m$). The corrected I-V characteristics of pm-junction using the band theory \cite{arumugam2020plasma} are given by the expression below
        \begin{equation}
	I=I_{es} \left[exp\frac{\left(V-V_P-\phi_m\right)}{T_e} \right]
	\end{equation}
Using these expressions for current the I-V characteristics of the pm-junction, the pn-junction, and the metal-semiconductor junction are obtained experimentally in the following section.

\section{Experiment}
The I-V characteristics of the pn-junction (homo-junction) and metal-semiconductor junction (hetero-junction, also known as Schottky diode) were obtained and compared with the pm-junction. The commercially available pn diode (1N4007 series)  and the metal-semiconductor junction diode (1N5822 series) are chosen for the experiment.  On the other hand, the pm-junction is made using a stainless steel and tungsten planar probe exposed to a DC glow discharge plasma \cite{arumugam2020plasma}. The experimental operating parameters have been optimised for stable plasma density and temperature, low cathode heating, etc. The experimental detail for each junction is described respectively in the following subsection.

\subsection{pn-junction and metal-semiconductor junction}
 The I-V characteristics of the pn-diode (1N4007) and Schottky diode (1N5822) were measured using dual-channel DC power supply (FALCON PS-302D, 0-30V/0-2A), $100$ $\Omega$ resistor, ammeter, and voltmeter to build the electric circuit. The current across the diodes was measured for applied voltage in the range of $-20V$ to $+20V$ to study the I-V characteristics of these two junctions.
 \begin{figure}[H]
		\includegraphics[width=\linewidth]{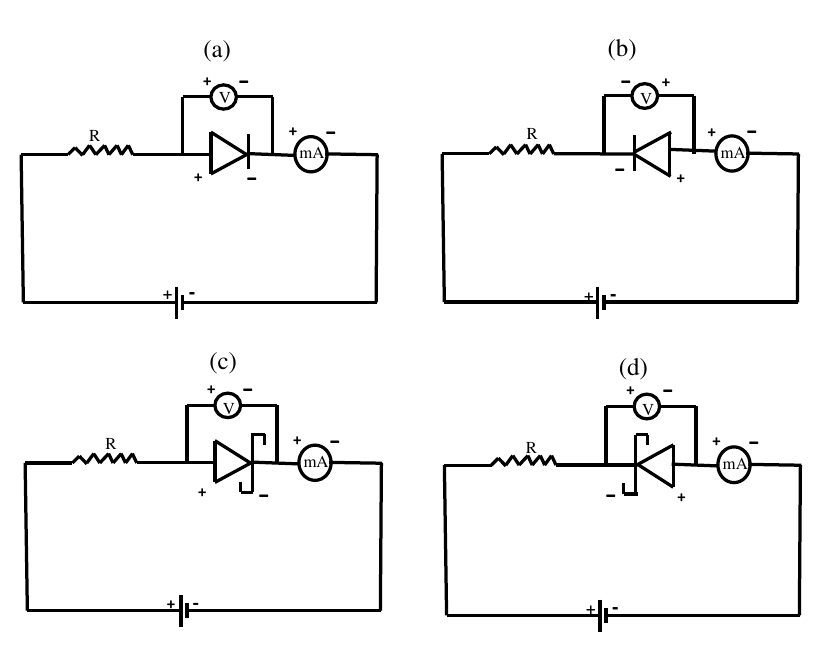} 
		\caption{Circuit diagram for pn diode (a) forward biased (b) reverse biased, and metal-semiconductor junction (c) forward biased (d) reverse biased.}
	\end{figure}
  
 \begin{figure}[H]
		\includegraphics[width=\linewidth]{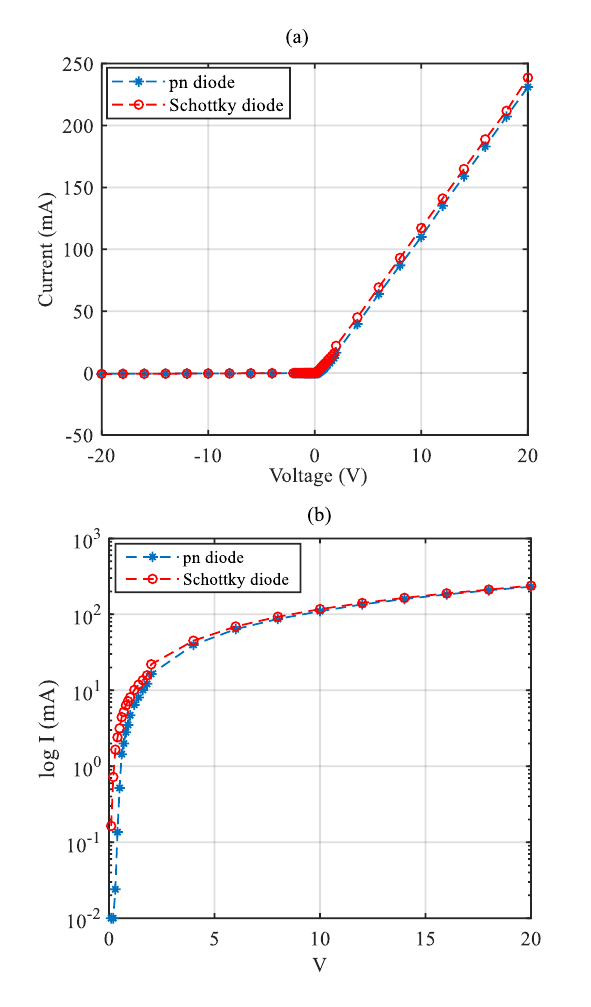} 
		\caption{I-V characteristics plot of (a) pn-diode (1N4007) and Schottky diode (1N5822) (b) semi-log plot of I-V characteristics shown in figure (a).}
	\end{figure}

 Fig. 2(a) shows the I-V characteristics of the pn-diode and the Schottky diode, and Fig. 2(b) is its semi-log plot for better visualization.  For a diode, when the forward bias voltage ($V$) is applied, the built-in potential barrier ($V_{bi}$) gets reduced to ($V_{bi}-V$), thus allowing more electrons to cross the barrier from the n-type region to the p-type region \cite{neeman2007semiconductor,streetman2000solid}. At low forward bias voltages, a minimal current flows up to the turn-on voltage, and beyond, it increases exponentially.  The magnified view of I-V characteristics is shown in Fig. 3(a) and 3(b), the turn-on voltages of the pn-diodes and Schottky diode are $0.5V$ and $0.24V$, respectively. These experimentally observed values of turn-on voltage match the standard values. The lower turn-on voltage for the Schottky diode is attributed to its smaller value of built-in potential barrier height than the pn-diode \cite{neeman2007semiconductor}. From $0V$ to $0.4V$, the pn-junction is resistive, and at $0.5V$, the exponential nature of the pn-diode arises, leading to a drop in the resistance. So, the turn-on voltage is around $0.5 V$, and further increasing to high voltage leads to a linear curve.
 \par
 In the reverse bias condition, the built-in potential barrier ($V_{bi}$) increased to ($V_{bi}+V$) with applied voltage $V$, thus forbidding electrons to cross the barrier \cite{neeman2007semiconductor,streetman2000solid}. This leads to a significantly less current for both. In reverse bias, the depletion layer’s width increases, and we get a very low current, of $\mu A$ values. The logarithmic scale of current against the linear voltage is shown in Fig. 3(b).
\begin{figure}[H]
		\includegraphics[width=\linewidth]{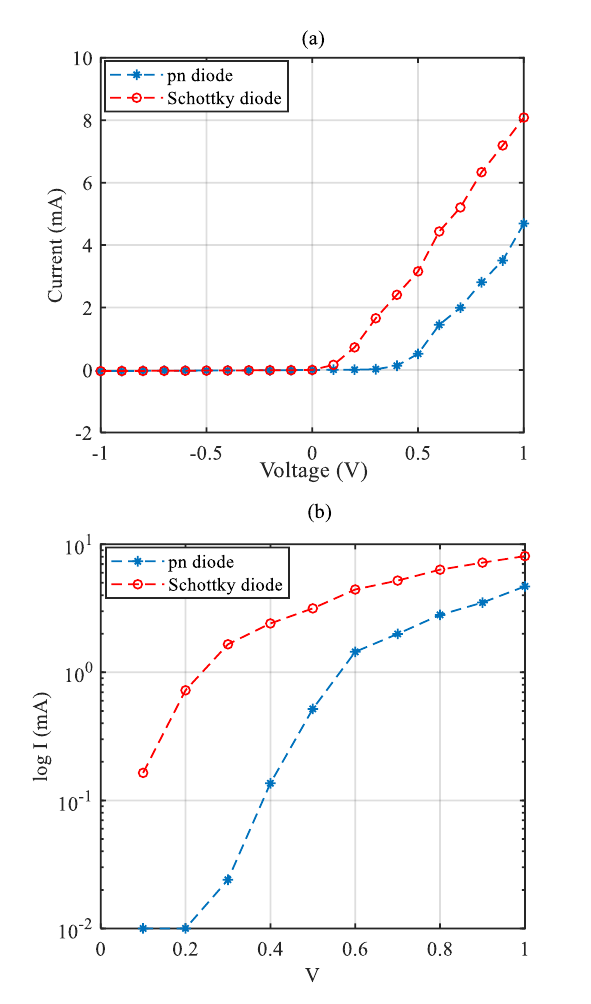} 
		\caption{Zoomed in I-V characteristics (a) pn-diode and Schottky diode (b) semi-log plot of (a).}
	\end{figure}

 \subsection{pm-junction}
	The experiment was done in a DC glow discharge plasma apparatus, as shown in Fig. 4. The chamber producing DC glow discharge plasma is made of stainless steel (SS304). The length and diameter of the chamber are $70$ $cm$ and $30$ $cm$. The discharge is produced between the grounded chamber and the cathode ($C$). Its diameter is $10$ $cm$ with $5$ $mm$ thickness, and it is made of stainless steel. Once the discharge is produced, plasma fills in the entire chamber. The pm-junction is made by exposing a tungsten (W) disc shape planar probe of diameter $2$ $cm$ and thickness $2$  $mm$.  Another pm-junction of identical size made up of stainless steel (SS) has been studied to study the effect of work function.  The probes making pm-junction were placed at a distance of $16$  $cm$ from the cathode to minimize the effect of the plasma density gradient\cite{lieberman1994principles}. Air is used as an operating gas, and the experiment was carried out at a pressure of $0.3$ $mbar$.\\
 \begin{figure}[H]
		\includegraphics[width=\linewidth]{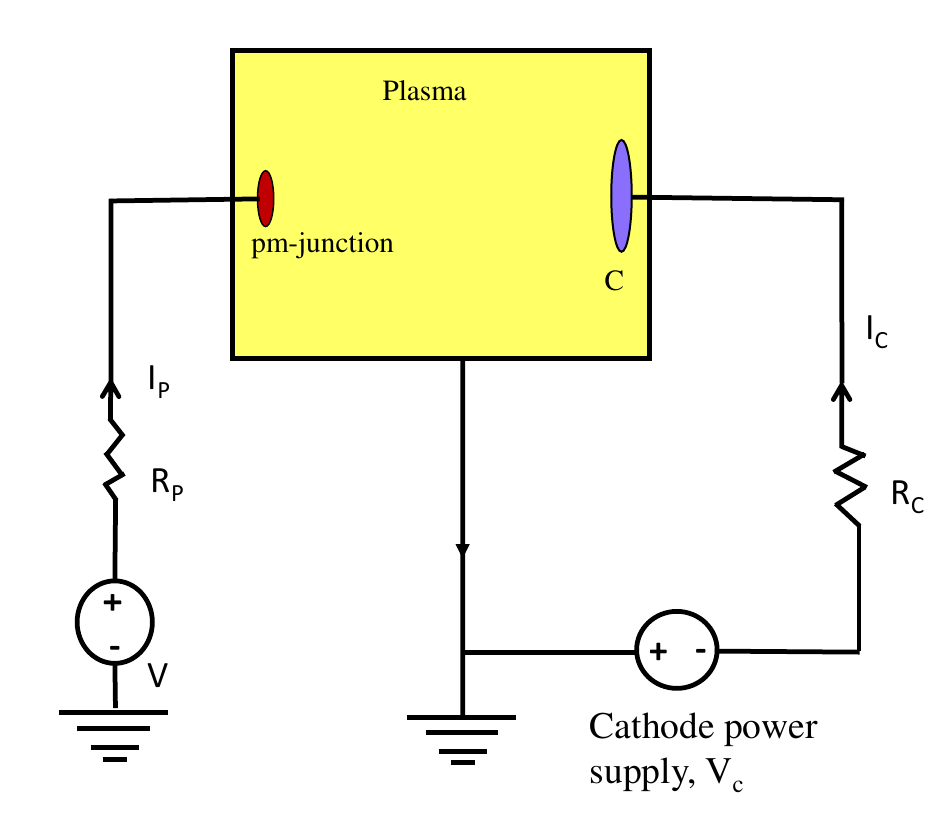} 
		\caption{Experimental setup of pm-junction.}
	\end{figure}
	As shown in Fig. 4, a DC power supply $(V_c)$ is used for biasing between the cathode and grounded chamber with a negative bias of $-500V$ to produce the plasma. A dual-channel DC power supply (FALCON PS-302D, 0-30V/0-2A) is used to apply voltage ($V$) on the probe to study the I-V characteristics. The voltage is applied in a narrow range of $-20V$ to $+20V$ to avoid any density perturbation and instabilities in the plasma\cite{alex2017triggering}.  The current ($I_{P}$) is measured through the resistor $R_{P}(100\Omega)$, connected in series with the power supplies. The I-V Characteristics of these two pm-junctions of two metals are recorded and further discussed in the next section. 
 
 \section{I-V Characteristics of pm-junction}
	Fig. 5 shows the variation in probe current on applying voltage ($V$). The I-V curve for stainless steel and tungsten pm-junction are shown for the range of $-20V$ to $+20V$. The turn-on voltages for both pm-junctions are $-7.0V$. This negative turn-on voltage is due to plasma electrons having higher energy, being free to move in the discharge volume, and readily reaching the probe surface. This results in drawing current even at $V=0V$. A negative voltage is required to stop them from reaching the probe surface. This is contrary to the electrons in the pn-junction and the Schottky diode, which need energy to overcome the potential barrier height to cross the junction.  However, the plot in Fig. 5(a) shows the difference in the current drawn by the W and SS probe, and the semilog plot in Fig. 5(b) gives a better visualisation of the I-V characteristics.   This difference is explained by expression (9), as the $(\phi_m)_W=4.55 eV$ for tungsten is higher than the $(\phi_m)_S=4.30 eV$ for SS, therefore the $I_P$ of tungsten is always less than SS.  \cite{arumugam2020plasma}. The current $I_P$ is in the range of $\sim \mu A$, and its profile is similar to the pn-diode and Schottky diode.
 \begin{figure}[H]
\includegraphics[width=\linewidth]{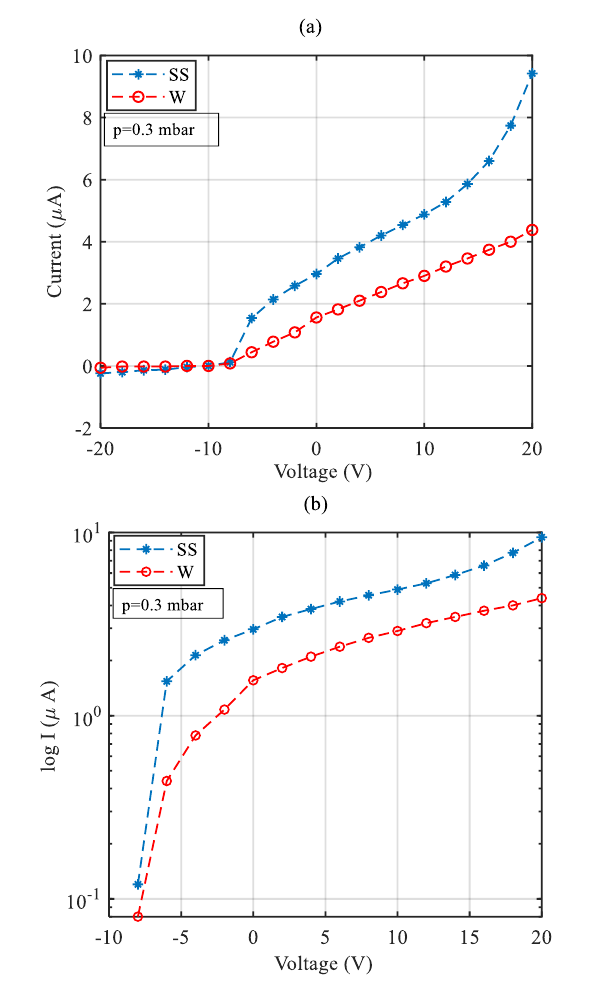}
\caption{I-V Characteristics plot ($-20V$ to $+20V$) of  (a) pm-junction-SS \& W and (b) semi-log plot of (a).}
\end{figure} 
 \par
The observed current range for pm-junction is in micro-amperes ($\mu A$), while for pn-diode and Schottky diode, the current is in the milli-amperes (mA) range.  Notwithstanding the different current ranges, the comparison of I-V characteristics is consistent. This large difference in current owes to the electron density $n$$ (m^{-3})$. In glow discharge plasma  $n$  $\sim$ $10^{16}$$m^{-3}$ \cite{lieberman1994principles};  while in metals $n$  $\sim$ $10^{28}m^{-3}$\cite{neeman2007semiconductor}. Even though the current can be attained in the range of  $m A$ for pm-junction for atmospheric pressure discharges or large probe sizes, the experimental operating parameters are optimised to ensure the repeatability of the experiment, stable discharge conditions, low electron density, and temperature fluctuations, and low heating effect  \cite{alex2017triggering,arumugam2020plasma,singh2023interplay}.    
\section{Discussion}
In this section, the observed I-V characteristics of pm-junction are analyzed, and its significance to gaseous electronics is discussed. Furthermore, we present current across pm-junction in analogy to pn-junction. Also, we propose the basic structure of the metal-oxide-plasma capacitor (MOP) structure in analogy to the metal-oxide-semiconductor (MOS) capacitor.
\subsection{Comparision of parameters}
The existing understanding of semiconductor physics explaining the formation of a potential barrier and its role in controlling the flow of electrons at a junction is the basis for the present study of plasma-metal junction. The relevant properties such as charge carrier density ($n$), Fermi energy for metals and semiconductors $(E_F=\frac{h^2}{8m}{\left[{\frac{3n}{\pi V}}\right]}^\frac{2}{3})$, and for plasma $E_F\approx eV_P$, Fermi temperature $(T_F=\frac{E_F} {k_B})$,  screening length for metals $(\lambda_F=\frac{\nu_F}{\omega_p})$, which is equivalent to Debye Length $(\lambda_D=\sqrt{\frac{KT\epsilon_0}{e^2n}})$ for plasma, average inter-particle distance  ($d=n^{-\frac{1}{3}}$), and the electrical conductivity ($\sigma$) of semiconductors, metals, and plasma for defining junction characteristics are listed in Table 1. The junction properties result from differences in the values of these electrical parameters for metals, semiconductors, and the glow discharge plasma. These parameters are useful for explaining and understanding the observed I-V curve. 
 \begin{table*}
		\centering
  \caption{Important electrical parameters of metals (W and SS), n-type and p-type semiconductors, and glow discharge plasma. The different parameters such as electron density $(n)$, Fermi energy ($E_F$), which is given by, Fermi temperature ($T_F$), screening length for metals $(\lambda_F)$, Debye length $(\lambda_D)$ for plasmas, average inter-particle distance $(d)$ \cite{manfredi2015solid}, and conductivity ($\sigma$)  are listed in the table.}
  
		\begin{tabular}{|m{5em}|m{6em}|m{6em}|m{7em}|m{7em}|}
			\hline	
			\multirow{2}{7em}{Parameters} &  \multicolumn{2}{c|}{Metals} & \multirow{2}{6em}{n-type/p-type Semiconductors
			\cite{neeman2007semiconductor}}& \multirow{2}{7em}{Glow discharge plasma \cite{chapmanbrian,uberoi1997introduction}}\\ 
			\cline{2-3}
			& $SS$ steel\cite{lide2004crc,uberoi1997introduction}& $W$
			\cite{arumugam2017effective,neeman2007semiconductor}& & \\
			\hline
			n $(m^{-3})$ & $17\times10^{28}$ & $13\times10^{28}$ & $10^{22}$ & $10^{16}$\\
			
			$E_F$ (eV)  & 11.1 & 6.4 & 0.9 & 30\\
			
			$T_F$ (K)& $12\times{10^4}$& $7\times{10^4}$& $1\times{10^4}$& $3.5\times{10^9}$\\
			
			$\lambda_F$, $\lambda_D$ (m)& $0.08 nm$& $0.07 nm$& $0.1\mu m$& $0.7\mu m$\\
			
			$d  (m)$& $0.8 nm$& $0.7nm$& $40 \mu m$& $0.2\mu m$\\
			
			$\sigma$ (S/m) & $1.42\times10^6$ & $18\times10^6$& $10$& $1.4\times10^3$\\
			\hline
		\end{tabular}
		\end{table*}
An interesting point to note is the electrical conductivity of metal $( 10^6 S/m)$and semiconductors $( 10 S/m)$, while for glow discharge plasma, the conductivity value $(\sim 10^3 S/m)$ is just intermediate between metal and semiconductor.  Nevertheless, the conductivity of plasma is a function of temperature and electron density and this investigation is limited to glow discharge plasma. The value of Fermi energy $E_F\approx 30$ $eV$ and Fermi temperature $T_F$ $(\sim 10^9 K)$ of the plasma is excessively high compared to the metals and semiconductors, which is obvious as the plasma state is of the highest energy state. Therefore, the pm-junction characteristics are expected to differ from the properties of the pn-diode and Schottky diode, and this is confirmed experimentally. The observed features of these three junctions are summarised in Table II. These junctions have different turn-on voltages. The turn-on voltage for the pn-junction and metal-semiconductor junction is $0.5V$ and $0.24V$, as shown in Figure 3. The turn-on voltage is a result of the typical barrier height, for the pn-junction is $ \sim 1eV$, and the metal-semiconductor junction is approx $\sim 0.52eV$ \cite{millman2010millman}. Contrary to this, for pm-junction, the turn-on voltage is $-7.0V$, much less than the normal diodes, shown in Fig. 5. This is because the energetic mobile plasma electrons rush to the metal surface \cite{chapmanbrian}, to prevent them from reaching the metal surface, a negative potential is required to be applied. Though the turn-on voltage for both pm-junctions, that is, for stainless steel and tungsten is the same  ($-7.0V$), the current depends on the work functions of metals, the higher the work function, the lower the current. This suggests that turn-on voltage only depends on plasma potential. For understanding, the schematic energy-band diagram, before and after the formation of the pm-junction, is shown in Fig. 6. 
\begin{table*}[t]
		\centering
  
  \caption{Junction parameters: Turn-on voltage ($V_{ON}$); Current at zero bias ($I_{0V}$); Current at 20V ($I_{20V}$)}
            
		\begin{tabular}{|m{5em}|m{5em}|m{9em}|c|m{3em}|}
			\hline	
			\multirow {2}{7em}{Parameters} & \multirow{2}{7em}{ pn-junction}& \multirow{2}{9em}{metal-semiconductor diode} & \multicolumn {2}{c|}{pm-junction(0.3 mbar)}\\
             \cline{4-5}
             & & & SS & W\\
			\hline
		\centering$V_{ON}$ & \centering0.5V & \centering0.24V & -7.0V & -7.0V\\
			
		\centering$I_{0V}$ & \centering0 & \centering0 & $2.96 \mu A$ & $1.56 \mu A$\\
			
		\centering$I_{20V}$ & \centering349.8 mA & \centering355.8 mA  & $ 631.4 \mu A$ & $8.64 \mu A$ \\
			\hline
		\end{tabular}
		
	\end{table*}
 
\begin{figure}[H]
		\includegraphics[width=\linewidth]{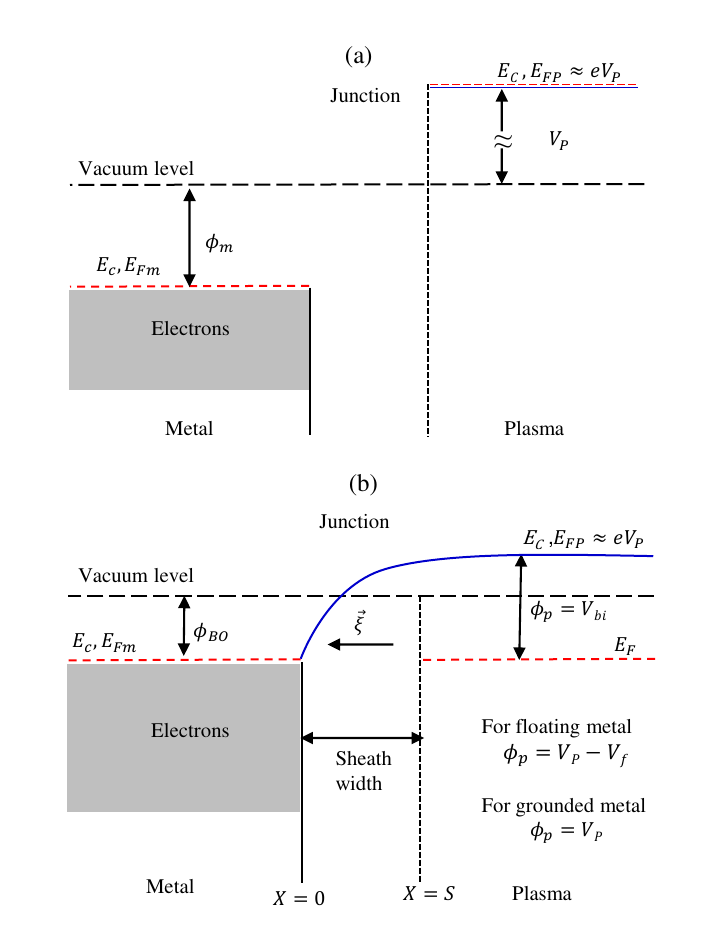} 
		\caption{Energy band diagram for pm-junction (a) Before junction formation (b) after junction formation. In the plasma region, the dotted red colour indicates the Fermi energy level ($E_F$), and the solid blue line indicates the top edge of the conduction band ($E_C$). Both the lines have coincided.}
	\end{figure}
Fig. 6 (a) shows the Fermi energy in the metal just coincides with the top edge of the conduction band, and the Fermi energy of plasma electrons above the vacuum level is the same as plasma potential. Fig. 6 (b) shows the formation of a pm-junction (which is generally known as a \textbf{\textit{sheath}} in plasma physics, and the junction width is called sheath thickness ($S$). The pm-junction potential $(\Phi_{p})$ is the difference between the plasma potential $(V_P)$ and the floating potential $(V_f)$ of the metal. In case the metal is grounded $V_f=0$, $\Phi_{p}=V_P$, which is the same as the built-in potential ($V_{bi}$) of the pn-junction\cite{arumugam2020plasma}. The plasma potential is always higher than any floating metal ( i.e., electrode) exposed to it, and thus the direction of the electric field ($\overrightarrow{\xi}$) at the junction is such that it prohibits loss of plasma electrons \cite{chen2012introduction}. Therefore, the pm-junction I-V curve is consistent with its energy-band diagram. For the negative applied voltage -20V to -10V, no current is drawn as plasma electrons do not have sufficient energy to overcome the applied potential, as shown in Fig. 5. On increasing the applied voltage above -7.0V, the junction current increases exponentially. As mentioned above, the negative turn-on voltage is a consequence of energetic plasma electrons.  Therefore, based on this investigation, two main conclusions are drawn. Firstly, the pm-junction potential barrier can be utilised to control the flow of electrons in a similar manner as it is done using the semiconductors junction. Secondly, the pm-junction turn-on voltage is negative, similar to the Gunn diode, and thus may be suitable for high-frequency applications.
A pn-junction is the fundamental basis of semiconductor devices, which describes that electrons could form potential barriers in some crystals. The flow of electrons can be controlled by manipulation of the barrier height.  This knowledge led to solid-state devices that substituted vacuum tubes and gaseous electronics for electrical operations, such as signal amplification. Many devices with various combinations of junctions were invented (such as n-p-n, p-n-p, p-n-p-n, etc.), and later integrated circuits were developed, leading to the information age's present era.  In this work, we have extended the junction physics for the junction of ionised gas, i.e., plasma and metal.  We believe that the present investigation on pm-junction may serve as a fundamental basis for gaseous electronics-based innovative applications such as plasma actuators, plasma shielding of fighter aircraft, plasma displays, and room temperature fast computing using the concept of chaos computing. 
\subsection{pm and pn-junction as a rectifier}
\par
The relevance of the present work is explained with the help of schematic diagrams in Fig. 7 and Fig. 8. In Fig. 7, the sinusoidal input signal for different voltages is applied across the pn-diode is shown. Fig. 7(a) shows that for a sinusoidal input signal having an amplitude of 0.5V to a pn-junction, there is no output, as the turn-on voltage for the pn-diode is +0.7V. For the 5.0V amplitude of the sinusoidal input wave, the output waveform is only a positive half cycle, and the pn-junction acts as a rectifier, as shown in Fig.7 (b). For the negative sinusoidal input wave of -10V in Fig. 7(c), no output is obtained.
\begin{figure}[H]
		\includegraphics[width=\linewidth]{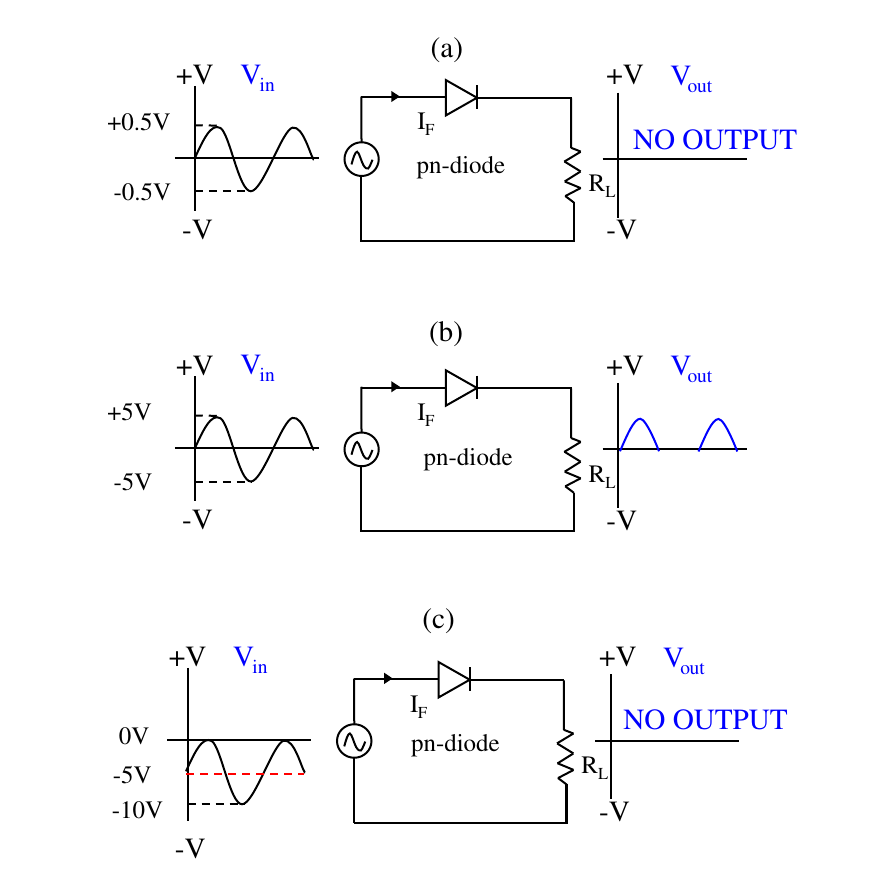} 
		\caption{Input and output waveforms of pn-diode, with a turn-on voltage +0.7V (a) no output as input is lower than turn-on voltage (b) half-wave rectifier as the negative half cycle is lower than turn-on voltage (c) no output as the input is lower than turn-on voltage}
	\end{figure} 
\par
The schematics shown in Fig. 8 present how a pm-junction with a turn-on voltage of -5.0V behaves differently than a pn-junction.  Fig. 8(a) shows that for a sinusoidal input signal having an amplitude of 0.5V to a pm-junction passes, as the turn-on voltage for the pm-diode is -5.0V. The 5.0V amplitude of the sinusoidal input signal is also able to pass across the pm-junction, as shown in Fig.8 (b). For the negative sinusoidal input wave of -10V as shown in Fig. 8(c), pm-junction rectifies it. Therefore, a pm-junction can be used as a rectifying as well as a non-rectifying junction.
\begin{figure}[H]
		\includegraphics[width=\linewidth]{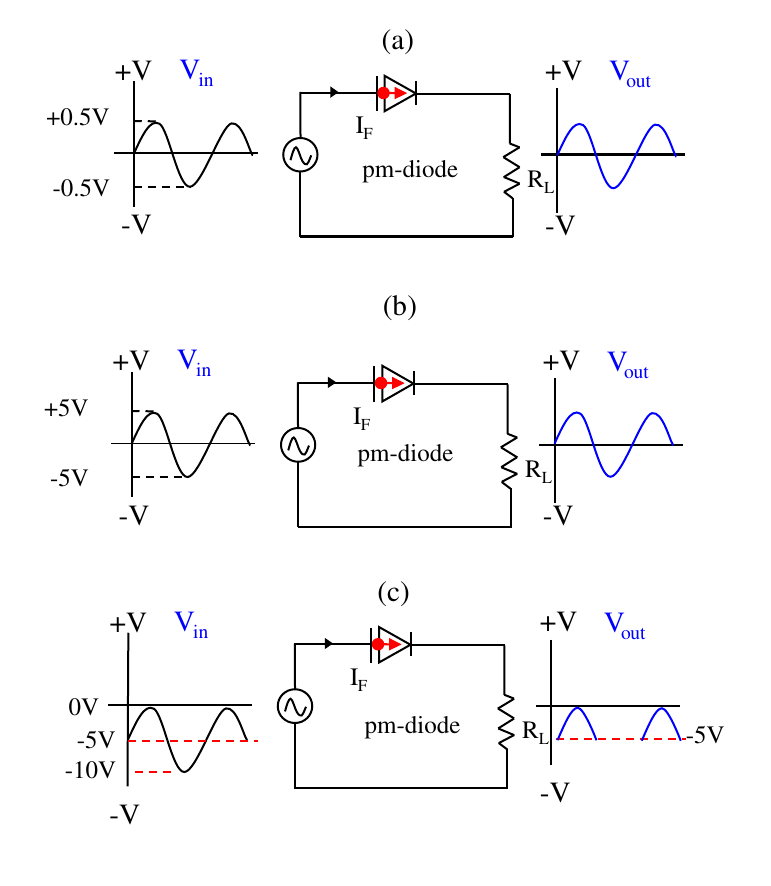} 
		\caption{Input and output waveforms of pm-diode, with a turn-on voltage -5V (a) Output is same as input, as input is higher than turn-on voltage (b) Output is same as input, as the input is higher than turn-on voltage (c) half-wave rectifier as half-cycle of the input is lower than turn-on voltage.}
	\end{figure}

\subsection{ Metal-Oxide-Plasma (MOP) capacitor: A new concept proposed  }
Though the present work is focused on the pm-junction, we propose how this understanding of the pm-junction can be extended to realise the metal-oxide-plasma junction analogous to the metal-oxide-semiconductor (MOS). The MOS and the proposed metal-oxide-plasma (MOP) structure are shown in Fig. 9(a) and 9(b). Fig. 9(b) shows a micro-discharge cell \cite{li2024review} in which discharge is sustained by the application of AC sustain voltage ($V_s$) on two parallel electrodes at the bottom of the cell. The oxide layer and the metal are layered similar to that of MOS. The energy band diagram of MOS is known, shown in Fig.9(c), and with the idea of an energy band diagram for pm-junction, the energy band diagram of MOP is drawn in Fig. 9 (d). The notations used in the energy band diagram are already given in Fig.6.  
\begin{figure}[H]
		\includegraphics[width=\linewidth]{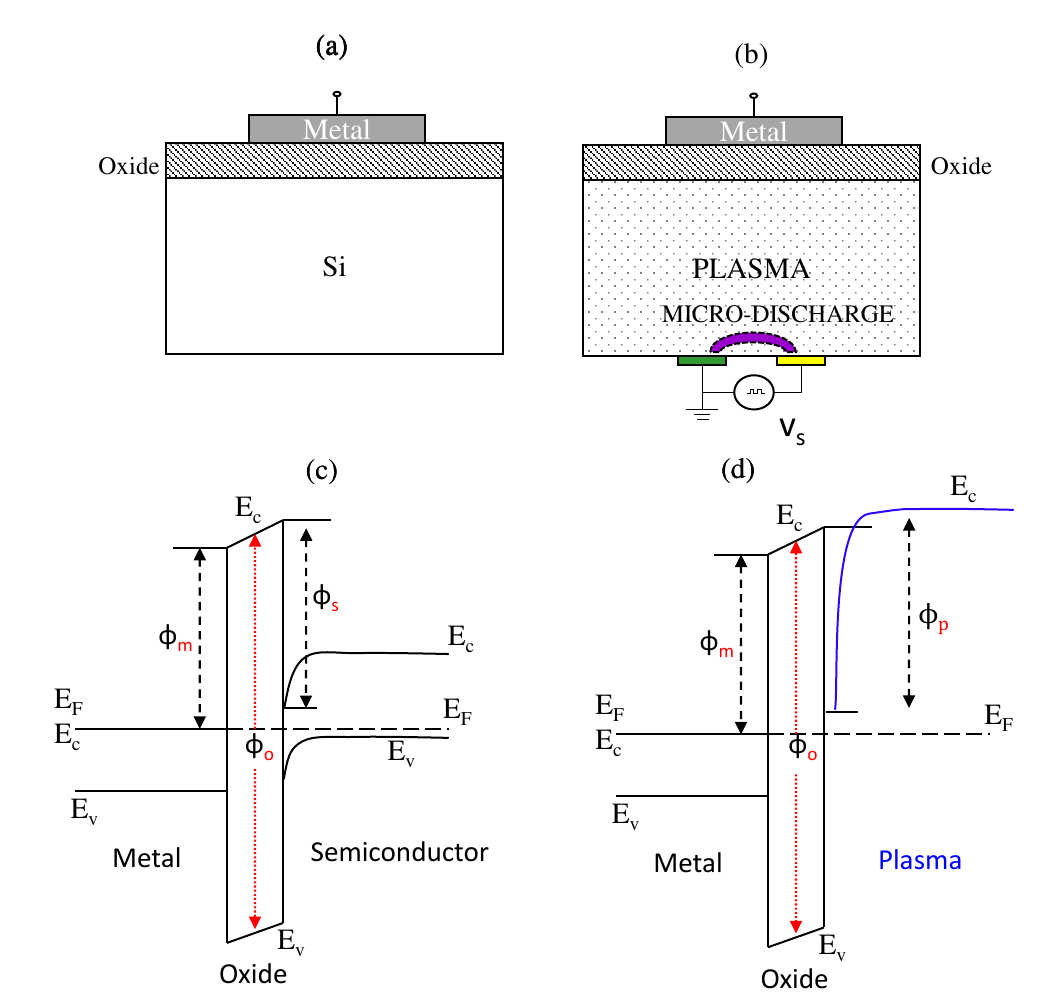} 
		\caption{The structure of (a) metal-oxide-semiconductor (MOS) (b) metal-oxide-plasma (MOP), and the energy band diagrams of (c) MOS, after contact (d) MOP, after contact. }
	\end{figure}
 \begin{figure}[H]
		\includegraphics[width=\linewidth]{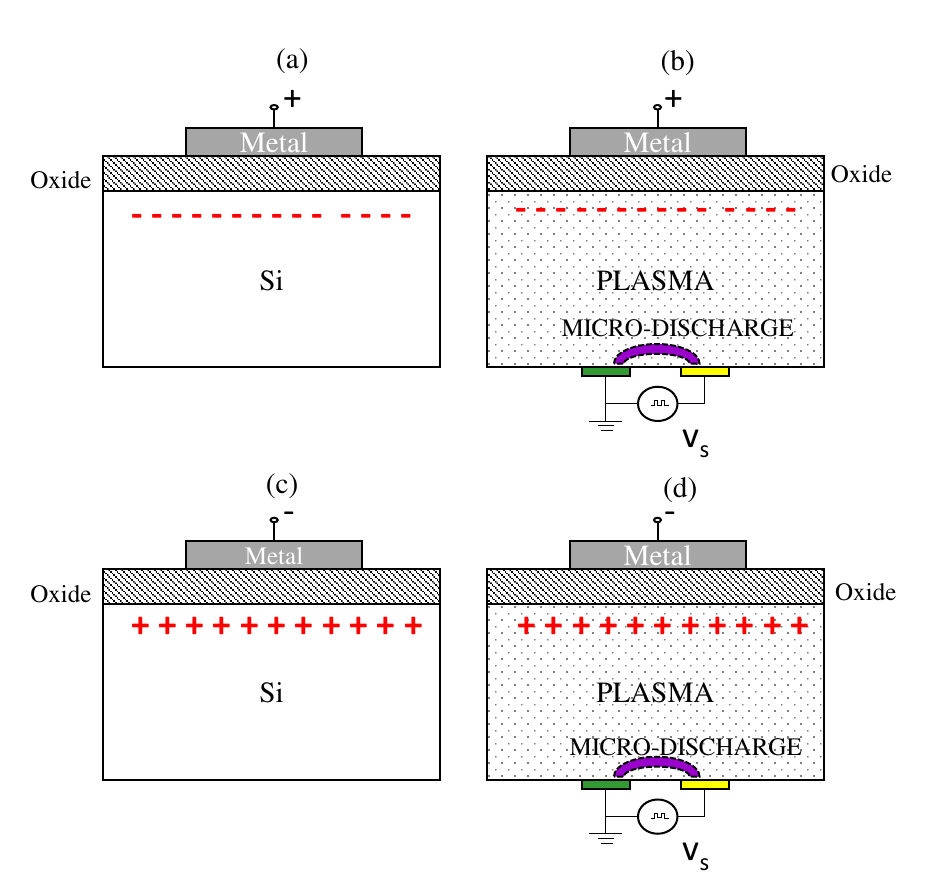} 
		\caption{The MOS and MOP capacitor is biased into surface accumulation (a) MOS with a positively biased gate (b) MOP with a positively biased gate (b) MOS with a negatively biased gate  (d) MOP with a negatively biased gate.}
	\end{figure}
The accumulation and depletion in the MOS capacitor with the applied biased voltage are well established. The schematic diagram shown in Fig. 10 (a) and Fig. 10 (b) describes that when a positive bias voltage is applied to the metal (i.e., GATE), both MOS and MOP accumulate electrons just below the oxide layer.  However, in case a negative bias voltage is applied to the metal (i.e., GATE), the electrons are pushed away from the oxide layer, exposing positive charges, as shown in Fig. 10 (c) and Fig. 10 (d).
\subsection{Similarities and differences between pm-junction and Vacuum tubes}
The present concept of pm-junction can be viewed as the applicability of the concept of junction physics, developed for understanding semiconductor junction, on GEDs. However, it is yet to be tested for vacuum tube electronics (VTE). In the case of vacuum tubes, the pm-junction will be modified in the sense that the quasi-neutral composition of ions and electrons i.e. the plasma, is replaced by an electron cloud.   In 1960, VTEs and GEDs were synonymous with each other, though the basic mechanisms differed. 
\par
I-V characteristics for both GEDs and VTEs have similar negative turn-on voltage and depend on electron density and temperature \cite{loeb2023basic}. However, the electron generation mechanism in VTEs is thermionic emission governed by the Richardson-Dushman equation \cite{richardson1916emission}. The current density (and thus electron density) depends on the temperature and work function of the material. However, in the case of a diode configuration and space-space charge limited condition current is expressed by Child-Langmuir law, which is true for cold plasmas also \cite{zhang2017100}.  Vacuum tubes used as switches made electronic computing possible for the first time. The 1946 ENIAC computer used 17,468 vacuum tubes and consumed 150 kW of power \cite{guarnieri2012age}. As a consequence of the high operating temperature, the VTE has lower device life and stability.
\par
On the other hand, in the case of gaseous discharge-based cold plasmas, the electron generation mechanism is avalanche ionization of gas led by the secondary electrons \cite{lieberman1994principles}. The electron density in the gas discharge produced is defined by the ion-induced secondary electron emission coefficient $(\gamma_e)$ of the cathode material, ionization coefficients of the gas (or gaseous mixture), and operating pressure. Therefore, gas discharge plasma provides control over a wide range of electron densities that can be produced depending on the application. Another interesting feature of pm-junction is its size can be scaled down to a few hundred nanometers and operate at sub-atmospheric pressure \cite{li2024review}. Self-processing abilities of plasmas, in addition to these electronic properties, may be applied suitably for fast computing at room temperature.
\section{Conclusions}
This investigation mainly aimed to compare the I-V characteristic of the pm-junction, a GED, to well-studied SSDs. We have chosen two commercially available SSDs, a pn-diode, and a Schottky diode. The junction characteristics of the pn-diode and the Schottky diode are well-known in semiconductor physics. In the experiment, the I-V curve is plotted for two pm-junctions, one of stainless steel (SS) and the other of tungsten (W), having different work functions. Its I-V curve is similar to the pn-junction and metal-semiconductor junction. The turn-on voltage for the pn-junction is $0.50 V$, higher than the turn-on Schottky barrier diode $0.24 V$, due to the larger potential barrier height. On the other hand, the plasma-metal junction has a negative turn-on voltage.  The turn-on voltage for both pm-junctions is  $-7.0V$, however, the junction with the higher work function draws a lower current. This suggests that turn-on voltage only depends on plasma potential and is consistent with the energy-band diagram of the pm-junction. 
\par
The present work demonstrates that the junction between metal and plasma can be utilized to control the flow of electrons in a similar manner to that of semiconductor junctions, and this concept of pm-junction may be further extended to plasma-insulator junction, plasma-semiconductor junction. It can act as a rectifying and non-rectifying contact, depending on the nature of the input signal. Further, we present a perspective application of this work for designing a GED-based alternative to SSD MOS device. Though a separate study for experimental demonstration of MOP characteristics is our high priority, it is beyond the scope of this letter.     
\section*{Acknowledgments}
The authors thank Pondicherry University for the University Fellowship. We are grateful to the Department of Electrical Engineering, Pondicherry University for the dual-channel DC power supply.

\bibliographystyle{unsrt}

\bibliography{manuscript}

\begin{thebibliography}{10}

\bibitem{neeman2007semiconductor}
DA~Neeman.
\newblock Semiconductor physics and devices, 2007.

\bibitem{jones1955gaseous}
Jones.F.
\newblock Gaseous electronics: Conference in new york.
\newblock {\em Nature}, 175:154--155, 1955.

\bibitem{loeb2023basic}
Leonard~B Loeb.
\newblock {\em Basic processes of gaseous electronics}.
\newblock Univ of California Press, 2023.

\bibitem{hirsh2012gaseous}
Merle Hirsh.
\newblock {\em Gaseous electronics}, volume~1.
\newblock Elsevier, 2012.

\bibitem{knill2010quantum}
Emanuel Knill.
\newblock Quantum computing.
\newblock {\em Nature}, 463(7280):441--443, 2010.

\bibitem{shinbrot1993using}
Troy Shinbrot, Celso Grebogi, James~A Yorke, and Edward Ott.
\newblock Using small perturbations to control chaos.
\newblock {\em nature}, 363(6428):411--417, 1993.

\bibitem{murali2003realization}
K~Murali, S~Sinha, and William~L Ditto.
\newblock Realization of the fundamental nor gate using a chaotic circuit.
\newblock {\em Physical Review E}, 68(1):016205, 2003.

\bibitem{bovzena2018non}
{\v{S}}ER{\'A} Bo{\v{z}}ena and {\v{S}}ER{\'Y} Michal.
\newblock Non-thermal plasma treatment as a new biotechnology in relation to seeds, dry fruits, and grains.
\newblock {\em Plasma Science and Technology}, 20(4):044012, 2018.

\bibitem{weltmann2016plasma}
KD~Weltmann and Th~Von~Woedtke.
\newblock Plasma medicine—current state of research and medical application.
\newblock {\em Plasma Physics and Controlled Fusion}, 59(1):014031, 2016.

\bibitem{kong2009plasma}
Michael~G Kong, G~Kroesen, G~Morfill, T~Nosenko, Toshimi Shimizu, J~Van~Dijk, and JL~Zimmermann.
\newblock Plasma medicine: an introductory review.
\newblock {\em new Journal of Physics}, 11(11):115012, 2009.

\bibitem{ishijima2013high}
T~Ishijima, K~Nosaka, Y~Tanaka, Y~Uesugi, Y~Goto, and H~Horibe.
\newblock A high-speed photoresist removal process using multibubble microwave plasma under a mixture of multiphase plasma environment.
\newblock {\em Applied Physics Letters}, 103(14), 2013.

\bibitem{mumtaz2023review}
S~Mumtaz, Rizwan Khan, Juie~Nahushkumar Rana, Rida Javed, Madeeha Iqbal, Eun~Ha Choi, and Ihn Han.
\newblock Review on the biomedical and environmental applications of nonthermal plasma.
\newblock {\em Catalysts}, 13(4):685, 2023.

\bibitem{hammer1999application}
Th~Hammer.
\newblock Application of plasma technology in environmental techniques.
\newblock {\em Contributions to Plasma Physics}, 39(5):441--462, 1999.

\bibitem{kogelschatz2004atmospheric}
Ulrich Kogelschatz.
\newblock Atmospheric-pressure plasma technology.
\newblock {\em Plasma Physics and Controlled Fusion}, 46(12B):B63, 2004.

\bibitem{alex2015order}
P~Alex, S~Arumugam, Kaliyamurthy Jayaprakash, and KS~Suraj.
\newblock Order--chaos--order--chaos transition and evolution of multiple anodic double layers in glow discharge plasma.
\newblock {\em Results in Physics}, 5:235--240, 2015.

\bibitem{alex2020coexistence}
P~Alex, M~Perumal, and S~K Sinha.
\newblock Coexistence of chaotic and complexity dynamics of fluctuations with long-range temporal correlations under typical condition for formation of multiple anodic double layers in dc glow discharge plasma.
\newblock {\em Nonlinear Dynamics}, 101(1):655--673, 2020.

\bibitem{jayaprakash2021doubly}
K~J~prakash, P~Alex, S~Arumugam, P~Murugesan, T~R Singh, and S~K Sinha.
\newblock Doubly forced anharmonic oscillator model for floating potential fluctuations in dc glow discharge plasma.
\newblock {\em PLA}, 410:127521, 2021.

\bibitem{singh2023interplay}
T~R Singh, P~Alex, and Suraj~Kumar Sinha.
\newblock Interplay between electron and ion plasma waves.
\newblock {\em Physics Letters A}, 477:128897, 2023.

\bibitem{rapp2013development}
J~Rapp, TM~Biewer, J~Canik, JBO Caughman, RH~Goulding, DL~Hillis, JD~Lore, and LW~Owen.
\newblock The development of plasma-material interaction facilities for the future of fusion technology.
\newblock {\em Fusion Science and Technology}, 64(2):237--244, 2013.

\bibitem{hess1990plasma}
D~W Hess.
\newblock Plasma--material interactions.
\newblock {\em Journal of Vacuum Science \& Technology A: Vacuum, Surfaces, and Films}, 8(3):1677--1684, 1990.

\bibitem{boeuf2003plasma}
JP~Boeuf.
\newblock Plasma display panels: physics, recent developments and key issues.
\newblock {\em Journal of physics D: Applied physics}, 36(6):R53, 2003.

\bibitem{roth2003aerodynamic}
J~Reece Roth.
\newblock Aerodynamic flow acceleration using paraelectric and peristaltic electrohydrodynamic effects of a one atmosphere uniform glow discharge plasma.
\newblock {\em Physics of plasmas}, 10(5):2117--2126, 2003.

\bibitem{arumugam2020plasma}
S~Arumugam, M~Perumal, KP~Anjana, SVM Satyanarayna, and Suraj~Kumar Sinha.
\newblock Plasma--metal junction.
\newblock {\em Physics of Plasmas}, 27(2), 2020.

\bibitem{bonitz2019towards}
M~Bonitz, A~Filinov, J~W Abraham, K~Balzer, H~K{\"a}hlert, E~Pehlke, F~X Bronold, et~al.
\newblock Towards an integrated modeling of the plasma-solid interface.
\newblock {\em Frontiers of Chemical Science and Engineering}, 13:201--237, 2019.

\bibitem{chapmanbrian}
Brian Chapman.
\newblock Brian chapman-glow discharge processes.
\newblock 1980.

\bibitem{lieberman1994principles}
Michael~A Lieberman and Allan~J Lichtenberg.
\newblock Principles of plasma discharges and materials processing.
\newblock {\em MRS Bulletin}, 30(12):899--901, 1994.

\bibitem{niu2022plasma}
Jiangqi Niu, Chayanaphat Chokradjaroen, Yasuyuki Sawada, Xiaoyang Wang, and Nagahiro Saito.
\newblock Plasma--solution junction for the formation of carbon material.
\newblock {\em Coatings}, 12(11):1607, 2022.

\bibitem{muthu2023On}
Muthu~Kumar.B Suraj Kumar~Sinha.
\newblock On the quantum treatment of the current through plasma-metal junction.
\newblock In {\em Proceedings of National Symposium on Gaseous Discharges (NSGD-2023), ISBN:978-81-8286-054-4}. T.R.Publications Pvt.Ltd, 2023.

\bibitem{neudeck1983pn}
G~W Neudeck.
\newblock {\em The PN junction diode}.
\newblock Addison-Wesley, 2nd ed, 1983.

\bibitem{dimitrijev2012principles}
Sima Dimitrijev.
\newblock {\em Principles of semiconductor devices}.
\newblock Oxford University Press, 2nd edition, 2012.

\bibitem{streetman2000solid}
Ben~G Streetman, Sanjay Banerjee, et~al.
\newblock {\em Solid state electronic devices}, volume~4.
\newblock Prentice hall New Jersey, 2000.

\bibitem{hershkowitz1989langmuir}
Noah Hershkowitz, O~Auciello, and DL~Flamm.
\newblock How langmuir probes work.
\newblock {\em Plasma diagnostics}, 1:113--183, 1989.

\bibitem{alex2017triggering}
P~Alex, Arumugham, and S~K Sinha.
\newblock Triggering of buneman instability and existence of multiple double layers in laboratory plasma.
\newblock {\em PLA}, 381(42):3652--3658, 2017.

\bibitem{manfredi2015solid}
Giovanni Manfredi and J{\'e}r{\^o}me Hurst.
\newblock Solid state plasmas.
\newblock {\em Plasma Physics and Controlled Fusion}, 57(5):054004, 2015.

\bibitem{uberoi1997introduction}
Chanchal Uberoi.
\newblock {\em Introduction to Unmagnetized Plasmas}.
\newblock PHI Learning Pvt. Ltd., 1997.

\bibitem{lide2004crc}
David~R Lide.
\newblock {\em CRC handbook of chemistry and physics}, volume~85.
\newblock CRC press, 2004.

\bibitem{arumugam2017effective}
Saravanan Arumugam, Prince Alex, and Suraj~Kumar Sinha.
\newblock Effective secondary electron emission coefficient in dc abnormal glow discharge plasmas.
\newblock {\em Physics of Plasmas}, 24(11), 2017.

\bibitem{millman2010millman}
Jacob Millman.
\newblock {\em Millman'S Integrated Electronics 2E}.
\newblock Tata McGraw-Hill education, 2010.

\bibitem{chen2012introduction}
Francis~F Chen.
\newblock {\em Introduction to plasma physics}.
\newblock Springer Science \& Business Media, 2012.

\bibitem{li2024review}
Yimeng Li, Lay~Kee Ang, Bing Xiao, Flyura Djurabekova, Yonghong Cheng, and Guodong Meng.
\newblock Review of electron emission and electrical breakdown in nanogaps.
\newblock {\em Physics of Plasmas}, 31(4), 2024.

\bibitem{richardson1916emission}
Owen~Willans Richardson.
\newblock {\em The emission of electricity from hot bodies}.
\newblock Longmans, Green and Company, 1916.

\bibitem{zhang2017100}
Peng Zhang, {\'A}g{\'u}st Valfells, LK~Ang, JW~Luginsland, and YY~Lau.
\newblock 100 years of the physics of diodes.
\newblock {\em Applied Physics Reviews}, 4(1), 2017.

\bibitem{guarnieri2012age}
Massimo Guarnieri.
\newblock The age of vacuum tubes: the conquest of analog communications [historical].
\newblock {\em IEEE Industrial Electronics Magazine}, 6(2):52--54, 2012.

\end{thebibliography}

\end{document}